\documentclass[a4paper]{article}
\usepackage{amssymb}

\usepackage{graphicx}
\usepackage{amsmath}

\begin{document}

\title{The separability of tripartite Gaussian state with amplification and
amplitude damping}
\author{Xiao-yu Chen \\
{\small {College of Information and Electronic Engineering, Zhejiang
Gongshang University, Hangzhou, 310018, China}}}
\date{}
\maketitle

\begin{abstract}
Tripartite three mode Gaussian state undergoes parametric amplification and
amplitude damping as well as thermal noise is studied. In the case of a
state totally symmetrically interacting with the environment, the time
dependent correlation matrix of the state in evolution is given. The
conditions for fully separability and fully entanglement of the final
tripartite three mode Gaussian state are worked out.


\end{abstract}

\section{Introduction}

Quantum entanglement of continuous variables (CV)\cite{Braunstein1} \cite
{Braunstein2} has received much attention recently, as shown by the
spectacular implementations of deterministic teleportation schemes\cite
{Vaidman} \cite{Furusawa} \cite{Takei} \cite{vanLoock1} \cite{Yonezawa},
quantum key distribution protocols \cite{Grosshans}, entanglement swapping
\cite{Takei} \cite{vanLoock2} \cite{Jia}, dense coding \cite{Ban}, quantum
state storage \cite{Julsgaard} and quantum computation \cite{Lloyd}
processes in quantum optical settings . Quantum information with CV in
general is mainly concerned with the family of Gaussian states, since these
comprise essentially all the experimentally realizable CV states. A
practical advantage of CV systems is the relative ease with which entangled
states can be generated in the laboratory \cite{Furusawa}. Former works are
mainly on the entanglement of bipartite system. The study of CV multipartite
entanglement which was initiated in \cite{vanLoock1} \cite{vanLoock2}, where
a scheme was suggested to create pure CV N-party entanglement using squeezed
light and N-1 beam splitters. In the practical situation, such a pure
multipartite entanglement state will evolve to a mixed state, due to the
decoherence. In all the multipartite CV states, tri-mode entangled state is
the simplest one, and a complete classification of tri-mode entanglement was
obtained, directly computable criterion that allows to determine to which
class a given state belongs\cite{Giedke} \cite{Chen1}. We in this paper will
investigate the separability of tripartite Gaussian state in presence of
amplitude damping, parametric amplification and noise which are symmetric
among the modes, based on our former works on the corresponding problem of
bipartite CV systems \cite{Chen} \cite{Chen2}.

\section{Time evolution of characteristic function}

The density matrix obeys the following master equation \cite{Kinsler} \cite
{Lindblad}\cite{Walls}
\begin{equation}
\frac{d\rho }{dt}=-\frac i\hbar [H,\rho ]+\mathcal{L}\rho .
\end{equation}
with the quadratic Hamiltonian
\begin{equation}
H=\hbar \sum_{jk}\frac i2(\eta _{jk}a_j^{\dagger }a_k^{\dagger }-\eta
_{jk}^{*}a_ja_k)
\end{equation}
where $\eta $ is a complex symmetric matrix. In the single-mode case, this
Hamiltonian describes two-photon downconversion from an undepleted
(classical) pump\cite{Walls}. The full multi-mode model describes
quasi-particle excitation in a BEC within the Bogoliubov approximation \cite
{Leggett}. This item represents the parametric amplifier. While the
amplitude damping is described by $\mathcal{L},$

\begin{equation}
\mathcal{L}\rho =\sum_j\frac{\Gamma _j}2\{(\overline{n}_j+1)L[a_j]\rho +%
\overline{n}_jL[a_j^{\dagger }]\rho  \nonumber
\end{equation}
where the Lindblad super-operator is defined as $L[\widehat{o}]\rho \equiv $
$2\widehat{o}\rho \widehat{o}^{\dagger }-\widehat{o}^{\dagger }\widehat{o}%
\rho -\rho \widehat{o}^{\dagger }\widehat{o}$.

We now transform the density operator master equation to the diffusion
equation of the characteristic function. Any quantum state can be
equivalently specified by its characteristic function. Every operator $%
\mathcal{A}\in \mathcal{B(H)}$ is completely determined by its
characteristic function $\chi _{\mathcal{A}}:=tr[\mathcal{AD}(\mu )]$ \cite
{Petz}, where $\mathcal{D}(\mu )=\exp (\mu a^{\dagger }-\mu ^{*}a)$ is the
displacement operator, with $\mu =[\mu _1,\mu _2,\cdots ,\mu _s]$ $%
,a=[a_1,a_2,\cdots ,a_s]^T$ and the total number of modes is $s.$ It follows
that $\mathcal{A}$ may be written in terms of $\chi _{\mathcal{A}}$ as \cite
{Perelomov}: $\mathcal{A}=\int [\prod_i\frac{d^2\mu _i}\pi ]\chi _{\mathcal{A%
}}(\mu )\mathcal{D}(-\mu ).$ The density matrix $\rho $ can be expressed
with its characteristic function $\chi $. $\chi =tr[\rho \mathcal{D}(\mu )]$
. Multiplying $\mathcal{D}(\mu )$ to the master equation then taking trace,
the master equation of density operator will be transformed to the diffusion
equation of the characteristic function. It should be noted that the complex
parameters $\mu _j$ are not a function of time, thus $\frac{\partial \chi }{%
\partial t}=tr[\frac{\partial \rho }{\partial t}\mathcal{D}(\mu )],$ the
parametric amplification part in the form of characteristic function will be
\cite{Walls}

\begin{equation}
\frac 12tr\{\sum_{jk}[\eta _{jk}a_j^{\dagger }a_k^{\dagger }-\eta
_{jk}^{*}a_ja_k,\rho ]D(\mu )\}=-\sum_{jk}(\eta _{jk}\mu _j^{*}\frac{%
\partial \chi }{\partial \mu _k}+\eta _{jk}^{*}\mu _j\frac{\partial \chi }{%
\partial \mu _k^{*}}).
\end{equation}

The master equation can be transformed to the diffusion equation of the
characteristic function, it is
\begin{eqnarray}
\frac{\partial \chi }{\partial t} &=&-\sum_{jk}(\eta _{jk}\mu _j^{*}\frac{%
\partial \chi }{\partial \mu _k}+\eta _{jk}^{*}\mu _j\frac{\partial \chi }{%
\partial \mu _k^{*}})  \label{wave} \\
&&-\frac 12\sum_j\Gamma _j\{\left| \mu _j\right| \frac{\partial \chi }{%
\partial \left| \mu _j\right| }+(2\overline{n}_j+1)\left| \mu _j\right|
^2)\chi \}.  \nonumber
\end{eqnarray}
Where we denote $\mu _j$ as $\left| \mu _j\right| e^{i\theta _j},$ and we
should take care about that the variables are $\mu _j$ and $\mu _j^{*}$ in
the amplification while they are $\left| \mu _j\right| $ $,\theta _j$ in the
damping.

\section{The parametric amplifier and the amplitude damping}

Suppose the solution to the diffusion equation is
\begin{eqnarray}
\chi (\mu ,\mu ^{*},t) &=&\chi (\nu ,\nu ^{*},0)\exp \{\frac 14(\nu ,-\nu
^{*})\left(
\begin{array}{ll}
\alpha & \beta ^{*} \\
\beta & \alpha ^{*}
\end{array}
\right) (\nu ^{*},-\nu )^T  \nonumber \\
&&-\frac 14(\mu ,-\mu ^{*})\left(
\begin{array}{ll}
\alpha & \beta ^{*} \\
\beta & \alpha ^{*}
\end{array}
\right) (\mu ^{*},-\mu )^T\},  \label{we0}
\end{eqnarray}
where $\nu =\mu M+\mu ^{*}N,$ with $M$ and $N$ being time varying matrices. $%
\alpha $ and $\beta $ are constant matrices and $\alpha ^{\dagger }=\alpha ,$
$\beta =\beta ^T$. Then $M$ and $N$ are the solutions of the following
matrix equations \cite{Chen}
\begin{eqnarray}
\frac{dM}{dt} &=&-\eta ^{*}N-\frac \Gamma 2M,  \label{we1} \\
\frac{dN}{dt} &=&-\eta M-\frac \Gamma 2N,  \label{we2}
\end{eqnarray}
where $\Gamma =diag\{\Gamma _1,\Gamma _2,\cdots ,\Gamma _s\}.$ While $\alpha
$ and $\beta $ are the solution of the following matrix equations
\begin{eqnarray}
2\eta \alpha +2\alpha ^{*}\eta -\Gamma \beta -\beta \Gamma +\Gamma w+w\Gamma
&=&0,  \label{we3} \\
\Gamma \alpha +\alpha \Gamma -2\eta ^{*}\beta -2\beta ^{*}\eta -\Gamma (2%
\overline{n}+1)-(2\overline{n}+1)\Gamma &=&0.  \label{we4}
\end{eqnarray}
with $\overline{n}=diag\{\overline{n}_1,\overline{n}_2,\cdots ,\overline{n}%
_s\}.$ The constant matrices $\alpha $ and $\beta $ can be worked out as the
solution of linear algebraic equations (\ref{we3}) and (\ref{we4}). What
left is to solve matrix equations (\ref{we1}) and (\ref{we2}). There are two
situations that the equations are solvable. The first case is that all the
modes undergo the same amplitude damping, that is $\Gamma _1=\Gamma
_2=\cdots =\Gamma _s$, thus $\Gamma =\Gamma _1\mathbf{I}_s.$ $\Gamma $
commutes with any matrix. Equations (\ref{we1}) and (\ref{we2}) have
solution
\begin{eqnarray}
M &=&e^{-\frac 12\Gamma t}\cosh ^{*}(\left| \eta \right| t),  \label{we5} \\
N &=&-e^{-\frac 12\Gamma t}\frac{\sinh (\left| \eta \right| t)}{\left| \eta
\right| }\eta .  \label{we6}
\end{eqnarray}
where the matrix cosh and sinh functions are defined as\cite{Corney}
\begin{eqnarray}
\cosh \left| \xi \right| &=&I+\frac 1{2!}\xi \xi ^{*}+\frac 1{4!}(\xi \xi
^{*})^2+\cdots , \\
\frac{\sinh \left| \xi \right| }{\left| \xi \right| }\xi &=&\xi +\frac
1{3!}\xi \xi ^{*}\xi +\frac 1{5!}(\xi \xi ^{*})^2\xi +\cdots .  \nonumber
\end{eqnarray}
The second case is that $\eta $ is a real matrix while the amplitude damping
can be different for each mode. The solution of equations (\ref{we1}) and (%
\ref{we2}) will be
\begin{eqnarray}
M &=&\frac 12[\exp (-\eta t-\frac{\Gamma t}2)+\exp (\eta t-\frac{\Gamma t}%
2)], \\
N &=&\frac 12[\exp (-\eta t-\frac{\Gamma t}2)-\exp (\eta t-\frac{\Gamma t}%
2)].
\end{eqnarray}

\section{The separability criterion of tripartite Gaussian state}

The separability problem of the three mode gaussian state was perfectly
solved\cite{Giedke}. The three mode gaussian states were classified as 5
different entangled classes\cite{Giedke}. But the states in this paper can
be classified as 3 different entangled classes: fully inseparable states,
biseparable states, fully separable states. Following the notation of Ref.
\cite{Giedke}, the CM $\alpha _{CM}$ now is replaced with $\gamma $, where $%
\gamma =2\alpha _{CM}$, the partial transposition is denoted as $\Lambda _j$
$(j=1,2,3$ is one of the three parties, here each party consists one mode$.$
If the canonical observables are arranged in the order of $%
x_1,x_2,x_3,p_1,p_2,p_3$, one has $\Lambda _1=diag\{1,-1,1,1,1,1\},$ $%
\Lambda _2=diag\{1,1,1,-1,1,1\}$ $\Lambda _3=diag\{1,1,1,1,1,-1\}.$ The
partially transposed CM will be $\widetilde{\gamma }_j=\Lambda _j\gamma
\Lambda _j$. Denote
\begin{equation}
J=\bigoplus_i\left[
\begin{array}{ll}
0 & -1 \\
1 & 0
\end{array}
\right]
\end{equation}
then criterion for fully inseparable state is

\begin{equation}
\widetilde{\gamma }_j\ngeq iJ,\text{ for all }j=1,2,3.
\end{equation}
Because of the symmetry of the 3ST state, the criterion can be simplified to
for example $\widetilde{\gamma }_A\ngeq iJ$ .

While for $\widetilde{\gamma }_j\geq iJ,($ $j=1,2,3),$ the state will be PPT
tri-mode state,and it can be biseparable or fully separable. The criterion
to distinguish the biseparable and fully separable states is \cite{Giedke}
as follow. The CM $\gamma $ of PPT tri-mode state can be written of as

\begin{equation}
\gamma =\left(
\begin{array}{ll}
A & C \\
C^T & B
\end{array}
\right) ,
\end{equation}
where $A$ is a $2\times 2$\ CM for the first mode, whereas $B$ is a $4\times
4$ CM for the other two modes. Define the matrices $K$ and $\widetilde{K}$ as

\begin{equation}
K\equiv A-C\frac 1{B-iJ}C^T,\text{ \qquad }\widetilde{K}\equiv A-C\frac 1{B-i%
\widetilde{J}}C^T,
\end{equation}
where $\widetilde{J}=J\oplus \left( -J\right) $ is the partially transposed $%
J$ for two modes.

Then the condition of the PPT tri-mode state being fully separable is that
if and only if there exists a point $(y,z)\in R^2$ fulfilling the following
inequality:
\begin{eqnarray}
\min \{\text{tr}K,\text{tr}\widetilde{K}\} &\geq &2x,  \label{wave5} \\
\det K+1+L^T\left(
\begin{array}{ll}
y, & z
\end{array}
\right) ^T &\geq &x\cdot \text{tr}K,  \label{wave6} \\
\det \widetilde{K}+1+\widetilde{L}^T\left(
\begin{array}{ll}
y, & z
\end{array}
\right) ^T &\geq &x\cdot \text{tr}\widetilde{K},  \label{wave7}
\end{eqnarray}
where $x=\sqrt{1+y^2+z^2},$ and $L=\left( u-w,2\text{Re}(v)\right) ,%
\widetilde{L}=\left( \widetilde{u}-\widetilde{w},2\text{Re}(\widetilde{v}%
)\right) $ if $K$ and $\widetilde{K}$ is written as

\begin{equation}
K=\left(
\begin{array}{ll}
u & v \\
v^{*} & w
\end{array}
\right) ,\text{ }\widetilde{K}=\left(
\begin{array}{ll}
\widetilde{u} & \widetilde{v} \\
\widetilde{v}^{*} & \widetilde{w}
\end{array}
\right) .
\end{equation}
Ineq.(\ref{wave5}) restricts $(y,z)$ to a circular disk $\mathcal{C}$, while
Ineq.(\ref{wave6}) and Ineq.(\ref{wave7}) describe ellipses $\mathcal{E}$
and $\mathcal{E}^{\prime }$ respectively. The existence of the point $(y,z)$
then turn out to be the intersection of the ellipses $\mathcal{E}$ and $%
\mathcal{E}^{\prime }$ and the circular disk $\mathcal{C}$.

The intersection of the ellipses $\mathcal{E}$ and $\mathcal{E}^{\prime }$
is a range in the $yz$ plane which is bounded by the elliptic curves $%
\partial \mathcal{E}$ and $\partial \mathcal{E}^{\prime }$. In the cases
considered in this paper, Re$(v)=0,$ Re$(\widetilde{v})=0,$ the two elliptic
curves $\partial \mathcal{E}$ and $\partial \mathcal{E}^{\prime }$ are
described by
\begin{eqnarray}
\det K+1+(u-w)y &=&(u+w)x,  \label{wave8} \\
\det \widetilde{K}+1+(\widetilde{u}-\widetilde{w})y &=&(\widetilde{u}+%
\widetilde{w})x,  \label{wave9}
\end{eqnarray}
$\partial \mathcal{E}$ and $\partial \mathcal{E}^{\prime }$ are centered at $%
y$ axis of $yz$ plane. The intersection of $\partial \mathcal{E}$ and $%
\partial \mathcal{E}^{\prime }$ is the solution of these two equations as
far as
\begin{equation}
x\geq \sqrt{1+y^2}.  \label{wave10}
\end{equation}
Thus the condition of the existence of $(\partial \mathcal{E})$ $\cap $ $%
(\partial \mathcal{E}^{\prime })$ is obtained.

Two situations of the intersection of the ellipses $\mathcal{E}$ and $%
\mathcal{E}^{\prime }$ and the circular disk $\mathcal{C}$ should be
considered. The first is $((\partial \mathcal{E})$ $\cap $ $(\partial
\mathcal{E}^{\prime }))\subseteq $ $\mathcal{C}$, in this case, fully
separability condition is determined by (\ref{wave8}) (\ref{wave9}) and (\ref
{wave10}). The second is $((\partial \mathcal{E})$ $\cap $ $(\partial
\mathcal{E}^{\prime }))$ $\varsubsetneq $ $\mathcal{C}$, in this case, we
should consider that if one of the tops of $\mathcal{E}\cap $ $\mathcal{E}%
^{\prime }$ is contained in $\mathcal{C}$ or not, the two tops are
determined by equations (\ref{wave8}) and (\ref{wave9}) separately by
setting $x=\sqrt{1+y^2}.$

\section{The symmetric amplification and damping of tripartite Gaussian state
}

We consider the totally symmetric amplification and amplitude damping among
all three modes, that is
\begin{equation}
\eta =\eta _0I+\eta _1S,
\end{equation}
with the matrix $S$ having its entries $S_{ij}=1$ for $i\neq j$ and $%
S_{ij}=0 $ for $i=j$ $(i,j=1,2,3);\Gamma _1=\Gamma _2=\Gamma _3,$ which are
redenoted as $\Gamma ,$ and $\overline{n}_1=\overline{n}_2=\overline{n}_3,$%
which are redenoted as $\overline{n}.$ For simplicity, we only consider the
case of real $\eta $. The matrix $\eta $ can be written as $\eta =U\zeta
U^{-1},$with $\zeta =diag\{\eta _0+2\eta _1,\eta _0-\eta _1,\eta _0-\eta
_1\}\equiv $ $\frac \Gamma 2diag\{\zeta _0,\zeta _1,\zeta _1\},$ and
\begin{equation}
U=\frac 1{\sqrt{6}}\left[
\begin{array}{lll}
\sqrt{2} & 0 & -2 \\
\sqrt{2} & \sqrt{3} & 1 \\
\sqrt{2} & -\sqrt{3} & 1
\end{array}
\right] .
\end{equation}
Then
\begin{eqnarray}
\cosh (\eta t) &=&Udiag\{\cosh (\zeta _0t^{\prime }),\cosh (\zeta
_1t^{\prime }),\cosh (\zeta _1t^{\prime })\}U^{-1}  \nonumber \\
&=&\frac I3[\cosh (\zeta _0t^{\prime })+2\cosh (\zeta _1t^{\prime })]+\frac
S3[\cosh (\zeta _0t^{\prime })-\cosh (\zeta _1t^{\prime })],
\end{eqnarray}
where $t^{\prime }=\frac \Gamma 2t,$ and there is the similar expression for
$\sinh (\eta t).$ The Eqs. (\ref{we3}) and (\ref{we4}) can be simplified in
the symmetric case and the solution is
\begin{eqnarray}
\alpha &=&(2\overline{n}+1)(\alpha _1I+\alpha _2S), \\
\beta &=&(2\overline{n}+1)(\beta _1I+\beta _2S),
\end{eqnarray}
with $\alpha _1=\frac 13(\frac 1{1-\zeta _0^2}+\frac 2{1-\zeta _1^2}),$ $%
\alpha _2=\frac 13(\frac 1{1-\zeta _0^2}-\frac 1{1-\zeta _1^2}),$ $\beta
_1=\frac 13(\frac{\zeta _0}{1-\zeta _0^2}+\frac{2\zeta _1}{1-\zeta _1^2}),$ $%
\beta _2=\frac 13(\frac{\zeta _0}{1-\zeta _0^2}-\frac{\zeta _1}{1-\zeta _1^2}%
).$

From (\ref{we0}), the complex CM at any time $t$ can be obtained in the real
symmetric amplification and symmetric damping case,
\begin{equation}
\gamma _c=\left[
\begin{array}{ll}
M & -N \\
-N & M
\end{array}
\right] [\gamma _c(0)-\left(
\begin{array}{ll}
\alpha & \beta \\
\beta & \alpha
\end{array}
\right) ]\left[
\begin{array}{ll}
M & -N \\
-N & M
\end{array}
\right] +\left(
\begin{array}{ll}
\alpha & \beta \\
\beta & \alpha
\end{array}
\right) ,
\end{equation}
where $M=e^{-\frac 12\Gamma t}\cosh (\eta t),$ $N=-e^{-\frac 12\Gamma
t}\sinh (\eta t).$ The real parameter CM can be reduced from the complex CM
and we have
\begin{equation}
\gamma =n^{\prime }\left[
\begin{array}{llllll}
a & 0 & c & 0 & c & 0 \\
0 & b & 0 & d & 0 & d \\
c & 0 & a & 0 & d & 0 \\
0 & d & 0 & b & 0 & d \\
c & 0 & c & 0 & a & 0 \\
0 & d & 0 & d & 0 & b
\end{array}
\right] ,
\end{equation}
with $n^{\prime }=2\overline{n}+1,$ and $a=\frac 13\exp [2(\zeta
_0-1)t^{\prime }](1-\frac{n^{\prime }}{1-\zeta _0})$ $+\frac 23\exp [2(\zeta
_1-1)t^{\prime }](1-\frac{n^{\prime }}{1-\zeta _1})$ $+\frac{n^{\prime }}%
3(\frac 1{1-\zeta _0}+\frac 2{1-\zeta _1}),$ $b=\frac 13\exp [-2(\zeta
_0+1)t^{\prime }](1-\frac{n^{\prime }}{1+\zeta _0})$ $+\frac 23\exp
[-2(\zeta _1+1)t^{\prime }](1-\frac{n^{\prime }}{1+\zeta _1})$ $+\frac{%
n^{\prime }}3(\frac 1{1+\zeta _0}+\frac 2{1+\zeta _1}),$ $c=\frac 13\exp
[2(\zeta _0-1)t^{\prime }](1-\frac{n^{\prime }}{1-\zeta _0})$ $-\frac 13\exp
[2(\zeta _1-1)t^{\prime }](1-\frac{n^{\prime }}{1-\zeta _1})$ $+\frac{%
n^{\prime }}3(\frac 1{1-\zeta _0}-\frac 1{1-\zeta _1}),$ $d=\frac 13\exp
[-2(\zeta _0+1)t^{\prime }](1-\frac{n^{\prime }}{1+\zeta _0})$ $-\frac
13\exp [-2(\zeta _1+1)t^{\prime }](1-\frac{n^{\prime }}{1+\zeta _1})$ $+%
\frac{n^{\prime }}3(\frac 1{1+\zeta _0}-\frac 1{1+\zeta _1}),$where the
vacuum initial state is assumed for simplicity.

\subsection{The fully separable conditions}

The condition (\ref{wave10}) of the existence of the intersection of the two
ellipses can be simplified to
\begin{equation}
-cd[(a-c)(b+2d)-1][(b-d)(a+2c)-1]\geq 0.
\end{equation}
In the limit of $t^{\prime }\rightarrow \infty ,$ we obtain $cd\leq 0$ for
all the parameters $\zeta _0$ and $\zeta _1,$ denote
\begin{eqnarray}
a^{\prime } &\equiv &a+2c=\exp [2(\zeta _0-1)t^{\prime }](1-\frac{n^{\prime }%
}{1-\zeta _0})+\frac{n^{\prime }}{1-\zeta _0}, \\
c^{\prime } &\equiv &a-c=\exp [2(\zeta _1-1)t^{\prime }](1-\frac{n^{\prime }%
}{1-\zeta _1})+\frac{n^{\prime }}{1-\zeta _1}, \\
b^{\prime } &\equiv &b+2d=\exp [-2(\zeta _0+1)t^{\prime }](1-\frac{n^{\prime
}}{1+\zeta _0})+\frac{n^{\prime }}{1+\zeta _0}, \\
d^{\prime } &\equiv &b-d=\exp [-2(\zeta _1+1)t^{\prime }](1-\frac{n^{\prime }%
}{1+\zeta _1})+\frac{n^{\prime }}{1+\zeta _1},
\end{eqnarray}
then the condition (\ref{wave10}) turns out to be
\begin{equation}
(a^{\prime }d^{\prime }-1)(c^{\prime }b^{\prime }-1)\geq 0.  \label{wq1}
\end{equation}

When the amplification is weaker the the damping, that is, $\max \{\left|
\zeta _0\right| ,\left| \zeta _1\right| \}<1.$ In the limit of $t^{\prime
}\rightarrow \infty ,$ all the time dependent terms in $a^{\prime
},b^{\prime },c^{\prime }$ and $d^{\prime }$ are damped to zeros. The final
CM tends to the residue CM, the condition of fully separability will be
\begin{eqnarray}
n^{\prime 2} &\geq &(1+\zeta _0)(1-\zeta _1),\text{ for }\zeta _0>\zeta _1,
\label{wave11} \\
n^{\prime 2} &\geq &(1-\zeta _0)(1+\zeta _1),\text{ for }\zeta _0<\zeta _1,
\label{wave12}
\end{eqnarray}
which can be rewritten as
\begin{eqnarray}
n^{\prime 2} &\geq &(1-\eta _0^{\prime }+\eta _1^{\prime })(1+\eta
_0^{\prime }+2\eta _1^{\prime }),\text{ for }\eta _1^{\prime }>0, \\
n^{\prime 2} &\geq &(1-\eta _0^{\prime }-2\eta _1^{\prime })(1+\eta
_0^{\prime }-\eta _1^{\prime }),\text{ for }\eta _1^{\prime }<0,
\end{eqnarray}
respectively, with $\eta _{0,1}^{\prime }=\frac 2\Gamma $ $\eta _{0,1}$.
These are the fully separable conditions of the residue states and shown in
Fig.1. Concerning the condition (\ref{wave5}), we consider the critical
situation of $n_0^{\prime 2}=(1+\zeta _0)(1-\zeta _1)$ for $\zeta _0>\zeta
_1 $, the condition (\ref{wave5}) reduces simply to $\zeta _0>\zeta _1,$
thus the state is fully separable in the critical situation. By physical
consideration, a state with $n^{\prime }>n_0^{\prime }$ then is fully
separable, because the state will be made more separable by adding the
noise. The same conclusion is true for the case of $\zeta _0<\zeta _1$.
Hence the fully separability conditions are given by (\ref{wave11}) and (\ref
{wave12}) in the situation of weak amplification.

Moreover, we can prove that (\ref{wave11}) and (\ref{wave12}) are also fully
separable conditions of the final state ($t^{\prime }\rightarrow \infty $)
for strong amplification situation. We provide the proof of one of the cases
here, the other cases can be followed with the same method. The case we
considered is $\zeta _0>1,\left| \zeta _1\right| <1,$ in the case we have $%
a^{\prime }\rightarrow +\infty ,b^{\prime }=\frac{n^{\prime }}{1+\zeta _0}%
,c^{\prime }=\frac{n^{\prime }}{1-\zeta _1},$ $d^{\prime }=\frac{n^{\prime }%
}{1+\zeta _1}>0,$ thus $a^{\prime }d^{\prime }>1$ and the condition (\ref
{wq1}) reduces to $b^{\prime }c^{\prime }\geq 1,$ which leads to (\ref
{wave11}). The condition (\ref{wave5}) can also be fulfilled for the
critical state in the strong amplification situation.
\begin{figure}[tbp]
\includegraphics[width=2.5in]{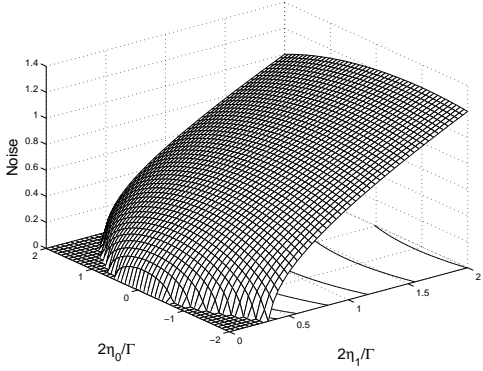}
\caption{The border of fully separable states and biseparable
states,characterized by the noise,the ratio of single mode amplification to
amplitude damping,the ratio of inter-mode amplification to amplitude
damping. The noise is the average noise $\overline{n}$.}
\end{figure}

\subsection{The biseparable conditions}

The biseparable condition $\widetilde{\gamma }_A\geq iJ$ can be simplified
to
\begin{equation}
1-(a^{\prime }b^{\prime }+8b^{\prime }c^{\prime }+8a^{\prime }d^{\prime
}+c^{\prime }d^{\prime })/9+a^{\prime }b^{\prime }c^{\prime }d^{\prime }\geq
0.  \label{wq2}
\end{equation}

In the weak amplification case, $\max \{\left| \zeta _0\right| ,\left| \zeta
_1\right| \}<1$. When $t^{\prime }\rightarrow \infty ,$ we have $a^{\prime }=%
\frac{n^{\prime }}{1-\zeta 0},$ $b^{\prime }=\frac{n^{\prime }}{1+\zeta _0},$
$c^{\prime }=\frac{n^{\prime }}{1-\zeta _1},$ $d^{\prime }=\frac{n^{\prime }%
}{1+\zeta _1}.$ The solution of Inequality (\ref{wq2}) is
\begin{equation}
n^{\prime 2}\geq 1-\frac 1{18}(\zeta _0^2+16\zeta _0\zeta _1+\zeta
_1^2)+\frac 1{18}\left| \zeta _0-\zeta _1\right| \sqrt{288+\zeta
_0^2+34\zeta _0\zeta _1+\zeta _1^2},
\end{equation}
which can also be written as
\begin{equation}
n^{\prime 2}\geq 1-\eta _0^{\prime 2}-\eta _0^{\prime }\eta _1^{\prime
}+\frac 32\eta _1^{\prime 2}+\frac 12\left| \eta _1^{\prime }\right| \sqrt{%
32+4\eta _0^{\prime 2}+4\eta _0^{\prime }\eta _1^{\prime }-7\eta _1^{\prime
2}},  \label{wq3}
\end{equation}

In the strong amplification case, we first consider the situation of $\zeta
_0>1,\left| \zeta _1\right| <1.$ When $t^{\prime }\rightarrow \infty ,$ we
have $a^{\prime }=\exp [2(\zeta _0-1)t^{\prime }](1+\frac{n^{\prime }}{\zeta
_0-1})+\frac{n^{\prime }}{1-\zeta _0}\rightarrow +\infty ,b^{\prime }=\frac{%
n^{\prime }}{1+\zeta _0},c^{\prime }=\frac{n^{\prime }}{1-\zeta _1},$ $%
d^{\prime }=\frac{n^{\prime }}{1+\zeta _1}.$ Denote $a^{\prime }=n^{\prime
}a_1,$ $b^{\prime }=n^{\prime }b_1,$ $c^{\prime }=n^{\prime }c_1,$ $%
d^{\prime }=n^{\prime }d_1,$ then Inequality (\ref{wq2}) is
\begin{equation}
1-(a_1b_1+8b_1c_1+8a_1d_1+c_1d_1)n^{\prime 2}/9+a_1b_1c_1d_1n^{\prime 4}\geq
0.  \label{wq4}
\end{equation}
The left hand side contain the linear and cubic terms of $n^{\prime }$
implicitly due to $a_1$. To solve the inequality, we just consider that $a_1$
does not depend on $n^{\prime }$ for the moment ( The solution to the
equality part of (\ref{wq4}) is a kind of iteration solution), the solution
to the quadratic inequality of $n^{\prime 2}$ can be obtained easily, at the
limitation of $a_1\rightarrow +\infty ,$ it is
\begin{eqnarray}
n^{\prime 2} &\geq &\frac 1{9c_1}(\frac 1{d_1}+\frac 8{b_1})  \nonumber \\
&=&(1-\zeta _1)(1+\zeta _0)-\frac 19(1-\zeta _1)(\zeta _0-\zeta _1).
\label{wq5}
\end{eqnarray}
The second situation we should consider is $\zeta _0>1,\zeta _1<-1,$ we have
$a_1\rightarrow +\infty ,d_1\rightarrow +\infty ,$ $b_1=\frac 1{1+\zeta _0},$
$c_1=\frac 1{1-\zeta _1}$ at $t^{\prime }\rightarrow \infty ,$ thus the
solution of Inequality (\ref{wq3}) reduces to
\begin{equation}
n^{\prime 2}\geq \frac 8{9b_1c_1}=\frac 89(1-\zeta _1)(1+\zeta _0).
\label{wq6}
\end{equation}
Similar results can be obtained for other domains of the parameters $\zeta
_0 $ and $\zeta _1,$ They are
\begin{eqnarray}
n^{\prime 2} &\geq &(1-\zeta _1)(1+\zeta _0)-\frac 19(1+\zeta _0)(\zeta
_0-\zeta _1),\text{ for}\left| \zeta _0\right| <1,\text{ }\zeta _1<-1;
\label{wq7} \\
n^{\prime 2} &\geq &(1+\zeta _1)(1-\zeta _0)-\frac 19(1-\zeta _0)(\zeta
_1-\zeta _0),\text{ for}\left| \zeta _0\right| <1,\text{ }\zeta _1>1;
\label{wq8} \\
n^{\prime 2} &\geq &(1+\zeta _1)(1-\zeta _0)-\frac 19(1+\zeta _1)(\zeta
_1-\zeta _0),\text{ for}\left| \zeta _1\right| <1,\text{ }\zeta _0<-1;
\label{wq9} \\
n^{\prime 2} &\geq &\frac 89(1+\zeta _1)(1-\zeta _0),\text{ for}\zeta _1>1,%
\text{ }\zeta _0<-1;  \label{wq10}
\end{eqnarray}
There are no restrictions to the $n^{\prime }$ in the domain of $\zeta
_1>1,\zeta _1>1$ and domain of $\zeta _0<-1,\zeta _0<-1,$ which means the
final states in these domains are always biseparable, in fact they are fully
separable according to (\ref{wave11}) and (\ref{wave12}). The synthesis of (%
\ref{wq3}) and (\ref{wq5})-(\ref{wq10}) is the biseparable condition for all
the parameters. We can transform the parameters to the single mode
amplification parameter $\eta _0$ and inter-mode amplification parameter $%
\eta _1$(appeared in the form of $\eta _0^{\prime }=2\eta _0/\Gamma ,\eta
_1^{\prime }=2\eta _1/\Gamma $), and the biseparable condition is shown in
Fig.2. A comparison of the figures as well as the formula show that the
fully separable and biseparable conditions are quite close with each other.
The detail of difference of the two is displayed in Fig.3 for the noiseless
situation.
\begin{figure}[tbp]
\includegraphics[width=2.5in]{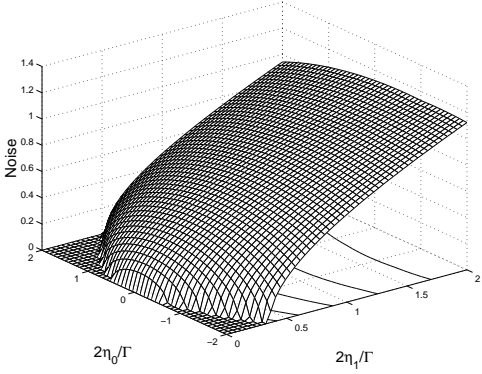}
\caption{The border of biseparable states and fully entangled
states,characterized by the noise,the ratio of single mode amplification to
amplitude damping,the ratio of inter-mode amplification to amplitude
damping. The noise is the average noise $\overline{n}$.}
\end{figure}

\begin{figure}[tbp]
\includegraphics[width=2.5in]{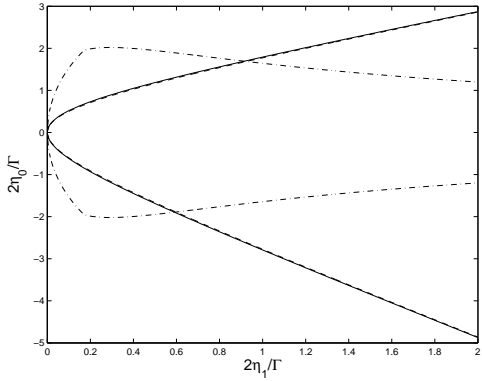}
\caption{The noiseless situation, the solid line for the border of fully
separable and biseparable states, the dashed line for the border of
biseparable and fully entangled states. The difference of the two curve is
very small, the detail of the difference is amplified by a factor of 100 and
shown by the dashdot line. }
\end{figure}

\section{Conclusion}

We consider the effect of amplification, amplitude damping and thermal noise
on tripartite three mode Gaussian state. The three modes are identical in
the initial preparation and in the later interaction with the environment
through amplification, damping and thermal noise. In such an assumption, the
analytical expression of fully separable and biseparable conditions for the
final tripartite three mode Gaussian state are obtained for both the weak
and strong amplification situation. In the weak amplification, no further
restriction on the initial state is required, besides the identical of the
three modes. While in the strong amplification case, vacuum initial state is
assumed for the simplicity of the description. The separability conditions
are characterized by the ratio of the single mode amplification parameter
with respect to the damping coefficient, the ratio of the inter-mode
amplification parameter with respect to the damping coefficient, and the
thermal noise. The biseparable condition and the fully separable condition
are very close with each other, make the domain of biseparable but not fully
separable states quite small. While both the domains of the genuine
entangled tripartite states and fully separable states are large enough. For
all the factor considered, the only factor to increase the entanglement is
the inter-mode amplification.

\section*{Acknowledgment}

Funding by the National Natural Science Foundation of China (under Grant No.
10575092), the Zhejiang Province Natural Science Foundation (under Grant No.
RC104265) are gratefully acknowledged.

\end{document}